\documentclass[12pt,preprint,showpacs,preprintnumbers]{revtex4}
\usepackage{amsmath}
\usepackage{graphicx}
\usepackage{dcolumn}
\usepackage{bm}
\usepackage{epsfig}
\usepackage[T1]{fontenc}
\usepackage{ae,aecompl}

\begin{document}

\title{Viscosity, heat conductivity and Prandtl number effects in
Rayleigh-Taylor Instability}
\author{Feng Chen$^1$\footnote{
Corresponding author. E-mail: shanshiwycf@163.com}, Aiguo
Xu$^{2,3}$\footnote{ Corresponding author. E-mail:
Xu\_Aiguo@iapcm.ac.cn}, Guangcai Zhang$^2$}
\affiliation{1, School of Aeronautics, Shan Dong Jiaotong University, Jinan 250357, China%
\\
2,National Key Laboratory of Computational Physics, Institute of
Applied Physics and Computational Mathematics, P. O. Box 8009-26,
Beijing 100088, China \\
3,Center for Applied Physics and Technology, MOE Key Center for
High Energy Density Physics Simulations, College of Engineering,
Peking University, Beijing 100871, China \\ }
\date{\today }

\begin{abstract}
Two-dimensional Rayleigh-Taylor(RT) instability problem is
simulated with a multiple-relaxation-time discrete Boltzmann model
with gravity term. The viscosity, heat conductivity and Prandtl
number effects are probed from the macroscopic and the
non-equilibrium views. In macro sense, both viscosity and heat
conduction show significant inhibitory effect in the
reacceleration stage, and the inhibition effect is mainly achieved
by inhibiting the development of Kelvin-Helmholtz instability.
Before this, the Prandtl number effect is not sensitive. Based on
the view of non-equilibrium, the viscosity, heat conductivity, and
Prandtl number effects on non-equilibrium manifestations, and the
correlation degrees between the non-uniformity and the
non-equilibrium strength in the complex flow are systematic
investigated.
\end{abstract}

 \maketitle

\section{Introduction}

The Rayleigh-Taylor (RT) instability\cite{rt1,rt2} occurs when a
heavy fluid lies above a lighter one in a gravitational field with
gravity pointing downward. The RT instability can be observed in a
wide range of astrophysical and atmospheric flows, and has great
significance in both fundamental research and practical
applications. Since the existence of sharp interfaces and their evolutions, the flow system is out of equilibrium.

Over the decades, many numerical methods have been developed to
simulate RT instability, such as flux-corrected transport
method\cite{ye1998}, level set method\cite{li1996}, front tracking
method\cite{Tryggvason2001}, marker-and-cell method\cite{Li2014},
smoothed particle hydrodynamics method\cite{Tang2004}, boundary
integral method\cite{Duchemin2005}, direct numerical simulations
\cite{Cook2001,Celani2006}, large-eddy
simulations\cite{Cabot2006}, and phase-field
method\cite{Celani2009}. The influences of different factors on
the evolution of RT instability have been studied more and more
deeply. R. Betti et al.\cite{Betti2006} investigated the effect of
vorticity accumulation on Ablative Rayleigh-Taylor Instability.
M.R.Gupta et al.\cite{Gupta2010} investigated the effect of
magnetic field, compressibility and density variation on the
nonlinear growth rate of RT instability. P.K. Sharma et
al.\cite{Sharma2010} analyzed the RT instability of two superposed
fluids taking the effect of small rotation, suspended dust
particles and surface tension. Rahul Banerjee et
al.\cite{Banerjee2011} investigated the combined effect of
viscosity and vorticity on the growth rate of the bubble
associated with single mode RT instability. To cite but a few. To
our knowledge, these numerical methods are based on the Euler or
Navier-Stokes equations, but Euler and Navier-Stokes models fall
short of describing the nonequilibrium effects.
Consequently, the rich and complex nonequilibrium effects in the RT flow system are rarely investigated.
 At the same time, the molecular dynamic simulations can present helpful information on the nonequilibrium state\cite{Kang-FOP2016}, but due to the limitation of compute capacity, the spatial and temporal scales it can access are far from large enough.

Besides the numerical methods mentioned above, the Lattice
Boltzmann (LB)
method\cite{SS,BSV,XGL1,XGL2,XGL3,ShanChen1,ShanChen2,goy,fwll,gs2013,xu2014,zxwsf}
provides an alternative efficient tool for simulating complex
fluid flows, and has been implemented in the RT instability
study\cite{nie1998,he1999a,he1999b,zhang2000,li2012,liu2013,liang2014,Sbragaglia2009,Scagliarini2010,Biferale2011}.
For instance, Nie et al. simulated the RT instability using a
lattice Boltzmann model for multicomponent fluid flows, and Guo et
al. investigated the effects of the Prandtl number on the mixing
process in RT instability of incompressible and miscible fluids
based on a double-distribution-function lattice Boltzmann method.
But up to now, in most of previous studies this LB method works as
a kind of new scheme to solve partial differential equations such
as the Euler equations and Navier-Stokes equations.

Recently, some scholars have re-positioned the method, and regard
it as a kind of new mesoscopic and coarse-grained kinetic model of
complex physical systems, which is juxtaposed with the traditional
hydrodynamic method and called as Discrete Boltzmann Method (DBM).
Compared with the first category, DBM possess more kinetic
information which is beyond the description of the Navier-Stokes,
and bring new physical insights into the physical system. The
first DBM description appeared in a review article published in
2012\cite{xu2012}. In the work, the authors pointed out how to
investigate both the Hydrodynamic Non-Equilibrium (HNE) and
Thermodynamic Non-Equilibrium (TNE) simultaneously in complex
flows via the DBM. Subsequently, DBM has been gradually extended
and applied to the combustion and detonation
system\cite{yxz2013,lxz2014,
xlz2015,xlz2015acta,lxz2016cnf,zxz2016cnf}, multiphase flow
system\cite{gxz2015} and fluid instability system\cite{lxz2014pre,
cxz2014, lai2015}. The finer physical structures of shock waves
revealed by DBM\cite{yxz2013,lxz2014,
xlz2015,xlz2015acta,lxz2016cnf,zxz2016cnf,lxz2014pre,cxz2014} have
been confirmed and suplemented by the results of non-equilibrium
molecular dynamics simulations\cite{kw2016}.

In this paper, we present a multiple-relaxation-time (MRT) DBM
with gravity. Two dimensional RT instability problem is simulated,
and the results are compared with those in previous studies. The
relaxation rates of the various kinetic moments due to particle
collisions may be adjusted more physically in the MRT version.
This overcomes some obvious deficiencies of the
Single-Relaxation-Time(SRT) version, such as a fixed Prandtl
number. Compared with previous studies on RT instability, the
viscosity, heat conductivity, and Prandtl number effects on
macro-dynamics and non-equilibrium manifestations are investigated
simultaneously in the DBM model. With the increase of viscosity or
heat conduction, various non-equilibrium components increase. When
the RT instability develops into the turbulent mixing stage, the
global average Thermodynamic NonEquilibrium (TNE) strength and
Non-Organized Energy Flux(NOEF) strength have a decrease. The
correlation degrees between density non-uniformity and the global
average TNE strength, temperature non-uniformity and the global
average NOEF strength, are numerically probed. And the simulation
results show that heat conduction plays a major role on the
correlation degree. The modeling of non-equilibrium feature is a
helpful and effective complement to the macroscopic description.
They two, together, provide new insights into complex flow
systems.

The following part of the paper is planned as follows. Section II
presents the MRT Discrete Boltzmann model with gravity. Systematic
numerical simulations of RT instability and non-equilibrium
characteristics are shown and analyzed in Section III. A brief
conclusion is given in Section IV.


\section{Description of the MRT DBM with gravity}

The MRT discrete Boltzmann
equation with gravity term read as follows%
\begin{equation}
\frac{\partial f_{i}}{\partial t}+v_{i\alpha }\frac{\partial f_{i}}{\partial
x_{\alpha }}=-\mathbf{M}_{il}^{-1}\hat{\mathbf{S}}_{lk}(\hat{f}_{k}-\hat{f}%
_{k}^{eq})-g_{\alpha }\frac{(v_{i\alpha }-u_{\alpha
})}{RT}f_{i}^{eq}\text{,}
\end{equation}%
where $\mathbf{v}_{i}$ is the discrete particle velocity,
$i=1$,$\ldots$ ,$N$, $N$\ is the number of discrete velocities.
The matrix $\hat{\mathbf{S}}=diag(s_{1},s_{2},\cdots ,s_{N})$ is
the diagonal relaxation matrix. $f_{i}$ and $\hat{f}_{i}$
($f_{i}^{eq}$ and $\hat{f}_{i}^{eq}$) are the particle
(equilibrium) distribution function in the velocity space and the
kinetic moment space respectively, the mapping between moment
space and velocity space is defined by the linear transformation
$M_{ij}$, i.e., $\hat{f}_{i}=M_{ij}f_{j}$,
$f_{i}=M_{ij}^{-1}\hat{f}_{j}$. $g_{\alpha }$ is the acceleration,
$u_{\alpha}$ is the macroscopic velocity, $T$ is the temperature.

Chapman-Enskog analysis indicates that it is independent of the
Discrete Velocity Model (DVM). Therefore, the choosing of DVM has
a high flexibility. Here, the following two-dimensional discrete
velocity model is used
\begin{align}
\left(v_{i1,}v_{i2}\right) =\left\{
\begin{array}{cc}
\mathbf{cyc}:c\left( \pm 1,0\right) , & \text{for }1\leq i\leq 4, \\
c\left( \pm 1,\pm 1\right) , & \text{for }5\leq i\leq 8, \\
\mathbf{cyc}:2c\left( \pm 1,0\right) , & \text{for }9\leq i\leq 12, \\
2c\left( \pm 1,\pm 1\right) , & \text{for }13\leq i\leq 16,%
\end{array}%
\right.  \label{dvm3}
\end{align}%
where \textbf{cyc} indicates the cyclic permutation,
$\eta_{i}=\eta _{0}$ for $i=1$, \ldots, $4$, and $\eta _{i}=0$,
for $i=5$, \ldots , $16$.
\begin{figure}[tbp]
\center\includegraphics*%
[bbllx=30pt,bblly=60pt,bburx=325pt,bbury=355pt,width=0.4\textwidth]{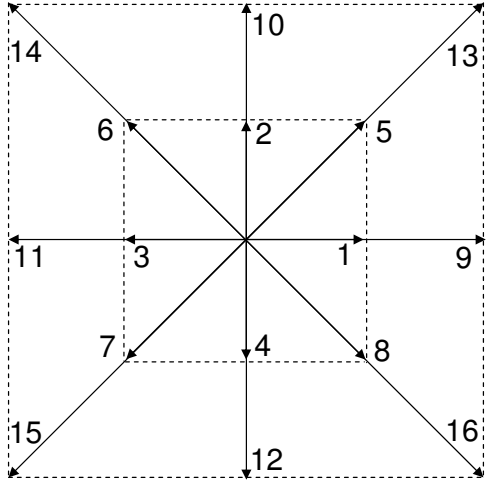%
}
\caption{Schematic of the discrete-velocity model.}
\end{figure}

Transformation matrix and the corresponding equilibrium
distribution functions in the Kinetic Moment Space are constructed
according
to the seven moment relations. Specifically, transformation matrix $\mathbf{M%
}=(m_{1},m_{2},\cdots ,m_{16})^{T}$, $m_{i}=(1,v_{ix},v_{iy},(v_{i\alpha
}^{2}+\eta _{i}^{2})/2,v_{ix}^{2},v_{ix}v_{iy},v_{iy}^{2},(v_{i\beta
}^{2}+\eta _{i}^{2})v_{ix}/2,(v_{i\beta }^{2}+\eta
_{i}^{2})v_{iy}/2,v_{ix}^{3},v_{ix}^{2}v_{iy},v_{ix}v_{iy}^{2},v_{iy}^{3},(v_{i\chi }^{2}+\eta _{i}^{2})v_{ix}^{2}/2,(v_{i\chi }^{2}+\eta _{i}^{2})v_{ix}v_{iy}/2,(v_{i\chi }^{2}+\eta _{i}^{2})v_{iy}^{2}/2)
$. The corresponding equilibrium distribution functions in KMS: $\hat{f}%
_{1}^{eq}=\rho \text{,}\qquad \hat{f}_{2}^{eq}=j_{x}\text{,}\qquad \hat{f}%
_{3}^{eq}=j_{y}\text{,}\qquad \hat{f}_{4}^{eq}=e\text{,}\qquad \hat{f}%
_{5}^{eq}=P+\rho u_{x}^{2}\text{,}\qquad \hat{f}_{6}^{eq}=\rho u_{x}u_{y}%
\text{,}\qquad \hat{f}_{7}^{eq}=P+\rho u_{y}^{2}\text{,}\qquad \hat{f}%
_{8}^{eq}=(e+P)u_{x}\text{,}\qquad \hat{f}_{9}^{eq}=(e+P)u_{y}\text{,}\qquad
\hat{f}_{10}^{eq}=\rho u_{x}(3T+u_{x}^{2})\text{,}\qquad $ $\hat{f}%
_{11}^{eq}=\rho u_{y}(T+u_{x}^{2})\text{,}\qquad \hat{f}_{12}^{eq}=\rho
u_{x}(T+u_{y}^{2})\text{,}\qquad \hat{f}_{13}^{eq}=\rho u_{y}(3T+u_{y}^{2})%
\text{,}\qquad \hat{f}_{14}^{eq}=(e+P)T+(e+2P)u_{x}^{2}\text{,}\qquad \hat{f}%
_{15}^{eq}=(e+2P)u_{x}u_{y}\text{,}\qquad \hat{f}%
_{16}^{eq}=(e+P)T+(e+2P)u_{y}^{2}$, where pressure $P=\rho RT$,
energy$\ e=b\rho RT/2+\rho u_{\alpha }^{2}/2$.

By using the Chapman-Enskog expansion on the two sides of the
discrete Boltzmann equation (see Appendix for details), the final
NS equations with gravity term for both compressible fluids and
incompressible fluids can be obtained:
\begin{subequations}
\begin{equation}
\frac{\partial \rho }{\partial t}+\frac{\partial (\rho u_{\alpha })}{%
\partial x_{\alpha }}=0\text{,}  \label{ns1}
\end{equation}%
\begin{equation}
\frac{\partial (\rho u_{\alpha })}{\partial t}+\frac{\partial \left( \rho
u_{\alpha }u_{\beta }\right) }{\partial x_{\beta }}+\frac{\partial P}{%
\partial x_{\alpha }}=\frac{\partial }{\partial x_{\beta }}[\mu (\frac{%
\partial u_{\alpha }}{\partial x_{\beta }}+\frac{\partial u_{\beta }}{%
\partial x_{\alpha }}-\frac{2}{b}\frac{\partial u_{\chi }}{\partial x_{\chi }%
}\delta _{\alpha \beta })]-\rho g_{\alpha }\text{,}  \label{ns2}
\end{equation}%
\begin{eqnarray}
\frac{\partial e}{\partial t}+\frac{\partial }{\partial x_{\alpha }}\left[
(e+P)u_{\alpha }\right] &=&\frac{\partial }{\partial x_{\beta }}[\lambda
\frac{\partial T}{\partial x_{\beta }}+\mu (\frac{\partial u_{\alpha }}{%
\partial x_{\beta }}+\frac{\partial u_{\beta }}{\partial x_{\alpha }}-\frac{2%
}{b}\frac{\partial u_{\chi }}{\partial x_{\chi }}\delta _{\alpha \beta
})u_{\alpha }]-\rho g_{\alpha }u_{\alpha }\text{,}  \label{ns3}
\end{eqnarray}%
where $\alpha, \beta, \chi=x, y$, the viscosity $\mu =\rho RT/s_{v}$, ($%
s_{v}=s_{5}=s_{6}=s_{7}$), the heat conductivity $\lambda =(\frac{b}{2}%
+1)\rho R^{2}T/s_{T}$, ($s_{T}=s_{8}=s_{9}$).


\section{Numerical Simulations}

\subsection{Performance on discontinuity}

\begin{figure}[tbp]
\center\includegraphics*%
[bbllx=15pt,bblly=15pt,bburx=310pt,bbury=240pt,width=0.5\textwidth]{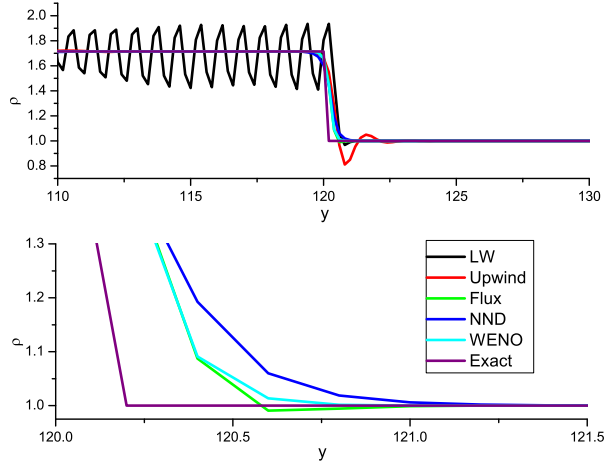}
\caption{Density profiles with various difference schemes at $t=12$.}
\end{figure}
In order to check the performance of difference scheme on
discontinuity, we construct this problem
\end{subequations}
\begin{equation}
\left\{
\begin{array}{cc}
(\rho ,u_{1},u_{2},T)=(1.71429,0.0,0.697217,1.26389), & y\leq L/2. \\
(\rho ,u_{1},u_{2},T)=(1.0,0.0,0.0,1.0), & L/2<y\leq L.%
\end{array}%
\right.
\end{equation}
$L$ is the length of computational domain. The physical quantities
on the two sides satisfy the Hugoniot relations, and specific heat
ratio $\gamma=1.4$. In the $y$ direction
$f_{i}=\mathbf{M}_{ij}^{-1}\hat{f}_{j}^{eq}$, and the macroscopic
quantities adopt the initial values. In the $x$ direction, the
periodic boundary condition is adopted. Fig.2 shows the simulation
results of density at time $t=12$ using different space
discretization schemes. The parameters
are $c=2$, $\eta_{0}=4$, $dx=dy=0.2$, $dt=10^{-4}$, $s_{i}=10^{4}$, $i=1$,$%
\ldots$ ,$16 $. The simulations with Lax-Wendroff scheme have
strong unphysical oscillations in the shocked region. The second
order upwind scheme results in unphysical `overshoot' phenomena at
the shock front. The simulation result with WENO scheme is much
more accurate, and decreases the unphysical oscillations at the
discontinuity.

\subsection{Macro-characteristics of Rayleigh-Taylor instability}

Numerical simulations of Rayleigh-Taylor instability are performed
in the section. The computational domain is a two-dimensional box
with height $H=80$ and width $W=20$, and the initial hydrostatic
unstable configuration is given by:
\begin{equation}
\left\{
\begin{array}{cc}
T_{0}(y)=T_{u};\rho _{0}(y)=\rho _{u}\exp (-g(y-y_{s})/T_{u}); & y\geq y_{s}
\\
T_{0}(y)=T_{b};\rho _{0}(y)=\rho _{b}\exp (-g(y-y_{s})/T_{b}); & y < y_{s} %
\label{rt}%
\end{array}%
\right.
\end{equation}
where $y_{s}=40+2 \cos(0.1\pi x)$ is the initial small perturbation at the
interface. To be at equilibrium, the same pressure at the interface should
be required
\begin{equation}
p_{0}=\rho _{u} T_{u}=\rho _{b} T_{b},  \label{rtp0}
\end{equation}
where $T_{u} < T_{b}$, $\rho _{u} > \rho _{b}$. In order to have a finite
width of the initial interface, all numerical experiments will been
performed by preparing the initial configuration plus a smooth interpolation
between the two half volumes. The initial temperature profile is therefore
chosen to be:
\begin{equation}
T_{0}(y)=(T_{u}+T_{b})/2+(T_{u}-T_{b})/2 \times tanh((y-y_{s})/w)
\end{equation}
where $w$ denotes the initial width of the interface. Initial density $\rho
_{0}(y)$ are then fixed by the initial settings (Eqs \eqref{rt}-\eqref{rtp0}%
) combined with the smoothed temperature profile. In the
simulation, the bottom condition is solid condition, the top
condition is free condition (that is to say, outflow condition),
and the left and right boundaries are periodic boundary
conditions. The fifth-order WENO scheme is used for space
discretization, while the time evolution is performed through the
third-order Runge-Kuta scheme.

In order to verify the validity of calculation, grid convergence study is
conducted in different grids, $Nx\times Ny = 100\times 400$ (grid I) and $%
Nx\times Ny = 200\times 800$ (grid II). The initial condition is $\rho
_{b}=1 $, $T_{b}=1.4$, $\rho _{u}=2.33333$, $T_{u}=0.6$, $g_{x}=0$, $%
g_{y}=0.005$, $w=0.8$, $\gamma=1.4$, the Atwood number is $A=0.4$.
Fig.3 shows the density and temperature distributions along the
line $x=5$ at time $t=200$, where $c=1$, $\eta _{0}=3 $,
$dt=10^{-3}$, all of the collision parameters are $10^{3}$. As one
can see, the agreement is good, and grid I is enough to simulate
the RT problem.
\begin{figure}[tbp]
\center\includegraphics*%
[bbllx=15pt,bblly=15pt,bburx=315pt,bbury=250pt,width=0.5\textwidth]{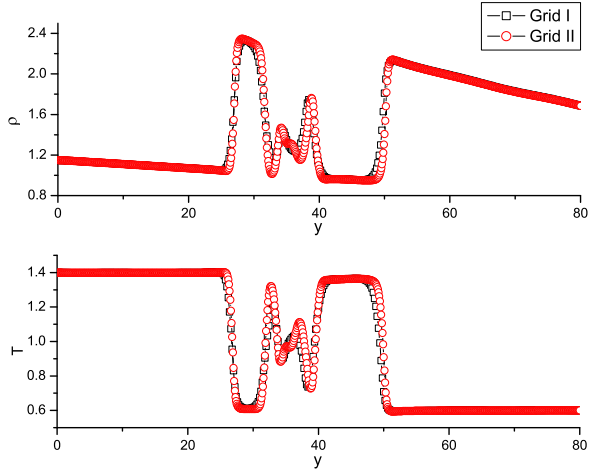}
\caption{Grid convergence study: the density and temperature
profiles at the line $x=5$ at time $t=200$.}
\end{figure}

Figure 4 shows the evolution of the fluid interface at time $t=0$, $100$, $%
200$, $300$, $400$. The bubble amplitude, spike amplitude, bubble
growth rate, and spike growth rate can be seen in Fig.5, and
represented by the black lines. When the amplitude of the
perturbation is much smaller than the wave length, the
perturbation of the fluid interface has an exponential growth. In
the spike formation stage, the heavy and light fluids gradually
penetrate into each other as time goes on, the light fluid rises
to form a bubble and the heavy fluid falls to generate a spike.
The interface becomes more acute and the growth rate is
approximately linearly increased. Subsequently, the
Kelvin-Helmholtz instability begins to develop and leads to the
accumulation of heavy fluid at the top of the spike. The interface
gradually becomes blunt, even eddy under certain conditions. The
spike growth rate  is reduced, and the bubble growth rate reaches
a constant velocity after a small attenuation. This is the
nonlinear stage. Taylor derived an empirical formula for the
constant velocity: $v_{b}=C\sqrt{AgW/2}$, where $C=0.32$. In the
simulation, the fitting constant speed of bubble is $0.05329$,
thus $C=0.3768$. The difference is due to the free condition at
the top. In a test of solid wall condition at the top, the fitting
constant velocity is $0.04622$, and $C=0.3268$. This agrees well
with Taylor and Layzer's results\cite{Layzer}. At a later time,
the extrusion from two sides leads to the formation of the
secondary spikes, and the growth rate increases again
(reacceleration stage). The shapes of the fluid interface in the
current study compare well with those in previous
studies\cite{hcz1999,lyzs2008}. The amplitude of spike is greater
than that of the bubble, and the ratio is changing with time.
After full development of the interface, the ratio is between
$1.5-1.7$, which is consistent with the numerical results of
Youngs\cite{youngs}.

\begin{figure}[tbp]
\center\includegraphics*%
[bbllx=5pt,bblly=200pt,bburx=596pt,bbury=575pt,width=0.75\textwidth]{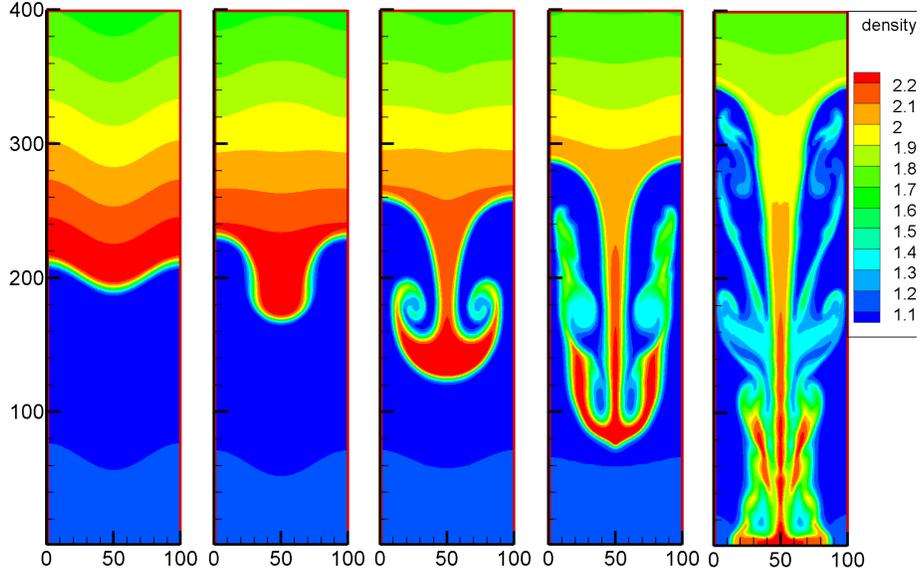%
}
\caption{Evolution of the fluid interface from a single mode perturbation.}
\end{figure}
\begin{figure}[tbp]
\center\includegraphics*%
[bbllx=15pt,bblly=15pt,bburx=305pt,bbury=230pt,width=0.64\textwidth]{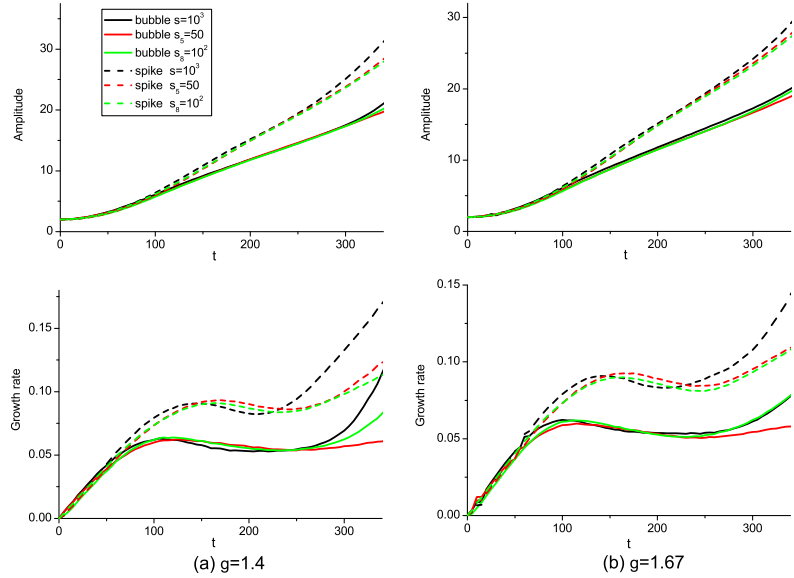%
} \caption{Amplitude and growth rate with different viscosity or
thermal conductivity.}
\end{figure}
\begin{figure}[tbp]
\center\includegraphics*%
[bbllx=15pt,bblly=15pt,bburx=320pt,bbury=125pt,width=0.75\textwidth]{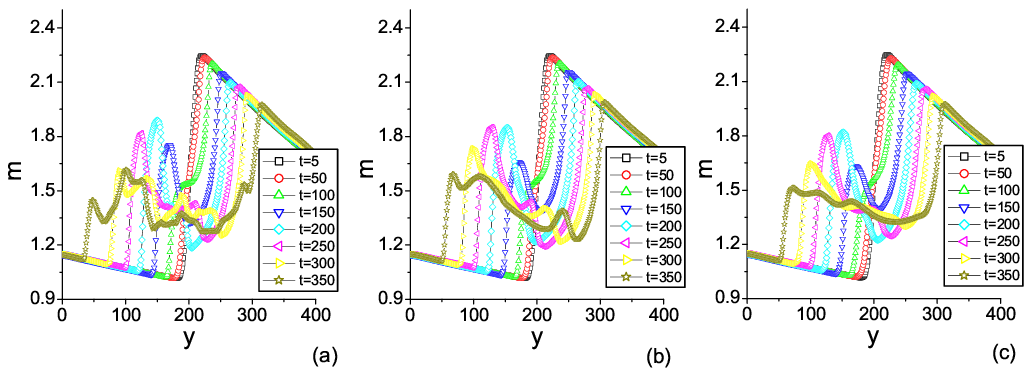%
} \caption{Evolution of the heavy material vertical distribution
curve.}
\end{figure}

Figure 6 shows the vertical distribution curve of heavy fluid
$m(y)$ at different times, which is defined as
\begin{equation}
m(y)=\sum_{ix=1}^{Nx}\rho(ix,iy)/NX.
\end{equation}
The occurrence and growth of the peak value of the heavy fluid
vertical distribution at time t=150, 200, represent the
accumulation of heavy fluid at the tip of the spike. Under the
extrusion action from two sides, the interface along the two
vortices is stretched, the peak value of heavy fluid vertical
distribution decreases gradually, and the distribution tends to be
approximate equilibrium.

The effects of viscosity and thermal conductivity on RT
instability are also shown in Figure 5, (a) $\gamma =1.4$, (b) $%
\gamma =1.667$. The black curves correspond to simulation results of $%
s=10^{3}$ ($Pr=1$), the red curves correspond to simulation results of $%
s_{v}=50$ (other collision parameters are $10^{3}$, $Pr=20$), and the green
curves correspond to $s_{T}=10^{2}$ (other collision parameters are $10^{3}$%
, $Pr=0.1$). Solid and dotted lines denote bubble and spike,
respectively. Before entering the reacceleration stage, the
effects of viscosity and thermal conductivity on RT instability
are negligible. At the reacceleration stage, both viscosity and
thermal conductivity show significant inhibitory effect. In Figure
7, we can find the explanations. (a), (b) and (c) correspond to
$Pr=20$, $Pr=1$ and $Pr=0.1$ respectively. With the decrease of
$s_{v}$ or $s_{T}$, the viscosity or thermal conductivity
increases, the complicated secondary vortices generated by the
Kelvin-Helmholtz instability are suppressed, and then the
evolution of RT instability is suppressed. That is to say, the
inhibition effect of viscosity and thermal conductivity on the RT
instability is mainly achieved by inhibiting the development of KH
instability in the RT instability.

\begin{figure}[tbp]
\center\includegraphics*
[bbllx=20pt,bblly=140pt,bburx=590pt,bbury=600pt,width=0.64\textwidth]{%
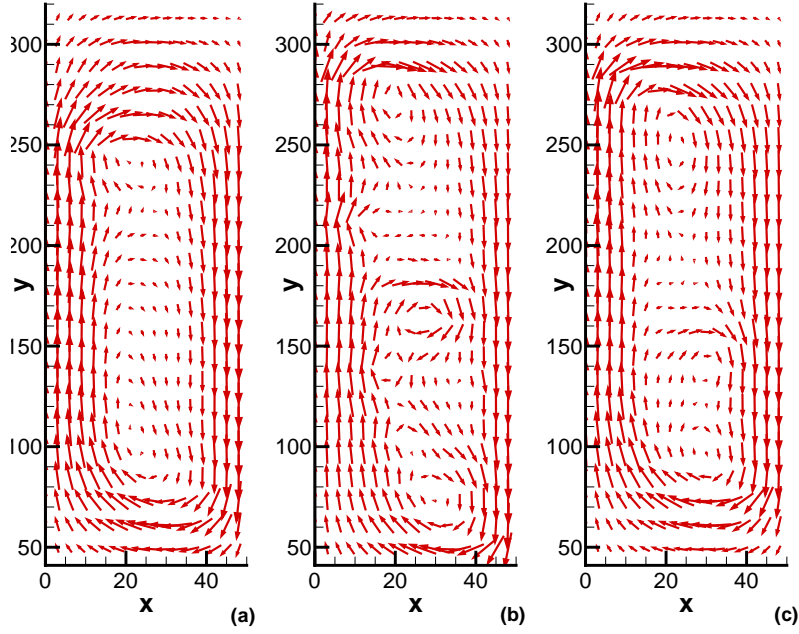} \caption{Velocity vector plots of RT instability
($\protect\gamma =1.4$) in the part of $[0,50]\times \lbrack
41,320]$ at time $t=350$, (a) $s_{v}=50$, (b) $s=10^{3}$, (c)
$s_{T}=10^{2}$.}
\end{figure}

\subsection{Non-equilibrium characteristic of Rayleigh--Taylor instability}

In the MRT model, the deviation from equilibrium can be defined as $\Delta
_{i}=\hat{f}_{i}-\hat{f}_{i}^{eq}=\mathbf{M}_{ij}(f_{j}-f_{j}^{eq})$. $%
\Delta _{i}$ contains the information of macroscopic flow velocity $%
u_{\alpha }$. Furthermore, we replace $v_{i\alpha }$ by $v_{i\alpha
}-u_{\alpha }$ in the transformation matrix $\mathbf{M}$, named $\mathbf{M}%
^{\ast }$ . $\Delta _{i}^{\ast }=\mathbf{M}_{ij}^{\ast
}(f_{j}-f_{j}^{eq})$ is only the manifestation of molecular
thermalmotion and does not contain the information of macroscopic
flow. In order to make the meaning of $\Delta _{i}^{\ast }$ more
clear, we introduce some symbols as $\Delta _{2xx}^{\ast }=\Delta
_{5}^{\ast }$, $\Delta _{2xy}^{\ast }=\Delta _{6}^{\ast }$,
$\Delta _{2yy}^{\ast }=\Delta _{7}^{\ast }$, $\Delta
_{(3,1)x}^{\ast }=\Delta _{8}^{\ast }$, $\Delta _{(3,1)y}^{\ast
}=\Delta _{9}^{\ast }$, $\Delta _{3xxx}^{\ast }=\Delta _{10}^{\ast
}$, $\Delta _{3xxy}^{\ast }=\Delta _{11}^{\ast }$, $\Delta
_{3xyy}^{\ast }=\Delta _{12}^{\ast }$, $\Delta _{3yyy}^{\ast
}=\Delta _{13}^{\ast }$, $\Delta _{(4,2)xx}^{\ast }=\Delta
_{14}^{\ast }$, $\Delta _{(4,2)xy}^{\ast }=\Delta _{15}^{\ast }$,
$\Delta _{(4,2)yy}^{\ast }=\Delta _{16}^{\ast }$. Here $\Delta
_{2xx}^{\ast }$ and $\Delta _{2yy}^{\ast }$ describe the
departures of the internal energies in the x and y degrees of
freedom from their average, $\Delta _{2xy}^{\ast }$ is concerned
with the
shear effects, $\Delta _{3xxx}^{\ast }$, $\Delta _{3xyy}^{\ast }$ and $%
\Delta _{(3,1)x}^{\ast }$ are related to the internal energy flow
caused by microscopic fluctuation in x direction, $\Delta
_{3xxy}^{\ast }$, $\Delta _{3yyy}^{\ast }$ and $\Delta
_{(3,1)y}^{\ast }$ are associated with the internal energy flow
caused by microscopic fluctuation in y direction. Compared with
the macroscopic equations, $\Delta_{2\alpha \beta}^{\ast }$ and
$\Delta _{(3,1)\alpha }^{\ast }$ correspond to the viscous stress
tensor in the momentum equation and the heat flux term in energy
equation, which are named as Non-Organized Momentum Flux (NOMF),
Non-Organized Energy Flux(NOEF), respectively\cite{ZYD}.

To provide a rough estimation of TNE, we follow the idea used in
refs.\cite{xlz2015}, and define a non-dimensional ``TNE strength''
function
\begin{equation*}
d(x,y)=\sqrt{\Delta _{2\alpha \beta}^{\ast 2}/T^2+\Delta
_{(3,1)\alpha}^{\ast 2}/T^3+\Delta _{3\alpha \beta \gamma}^{\ast
2}/T^3+\Delta _{(4,2)\alpha}^{\ast 2}/T^4}
\end{equation*}%
where $d=0$ in the thermodynamic equilibrium, and $d>0$ in the
thermodynamic nonequilibrium state. $D_{TNE}=\overline{d}$ is the
global
average TNE strength. Then we define $D_{2}=\overline{\sqrt{%
\Delta _{2\alpha \beta }^{\ast 2}}}$ and $D_{(3,1)}=\overline{\sqrt{%
\Delta _{(3,1)\alpha }^{\ast 2}}}$, $D_{2}$ and $D_{(3,1)}$ are
the global average NOMF strength and NOEF strength.
Correspondently, a macroscopic non-uniformity function is defined
\begin{equation*}
\delta W(x,y)=\sqrt{\overline{(W-\overline{W })^{2}}}
\end{equation*}
where $W=(\rho, U, T)$ denotes the macroscopic distribution,
$\overline{W}$ is the average value of a small cell around the
point $(x,y)$.

\begin{figure}[tbp]
\center\includegraphics*%
[bbllx=10pt,bblly=10pt,bburx=315pt,bbury=210pt,width=0.64\textwidth]{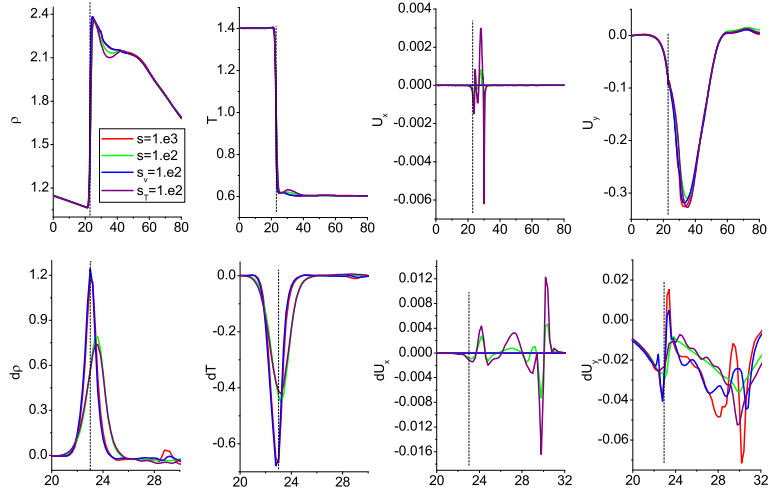%
}
\caption{Physical quantities and their gradients in the line $x=10$ at time $%
t=225$.}
\end{figure}
\begin{figure}[tbp]
\center\includegraphics*
[bbllx=12pt,bblly=15pt,bburx=234pt,bbury=205pt,width=0.85\textwidth]{%
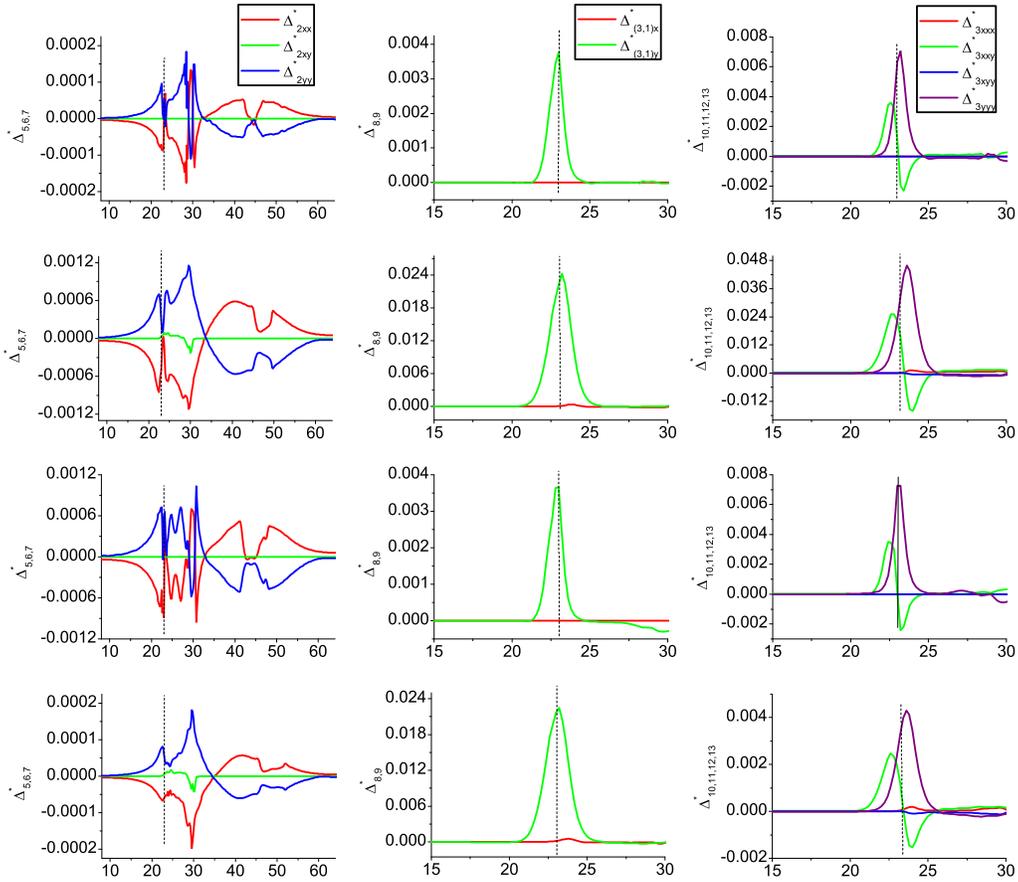} \caption{Non-equilibrium characteristics in the line
$x=10$ at time $t=225$ in four cases}
\end{figure}

Here we first give some results of $\Delta_{i}^{\ast}$ in the
evolution of RT instability. The initial physical quantities
($\rho, T, u_{x}, u_{y}$) are given the same values as those in
Fig 4. Figure 8 shows the simulation results
of physical quantities and their gradients in the line $x=10$ at time $t=225$%
. Fig. 9 shows the non-equilibrium characteristics of RT
instability with different viscosity or heat conduction. The first
line corresponds to $s=10^3$ (case I), the second
line corresponds to $s=10^2$ (case II), the third line corresponds to $%
s_{v}=10^2$ (case III), and the fourth line corresponds to
$s_{T}=10^2$ (other collision parameters are $10^3$, case IV). A
vertical dashed line is plotted in each panel to guide the eye for
the peak of spike. From Figs. 8 and 9, we can get the following
information.

1) $\Delta_{2xx,2yy,(3,1)y,3xxy,3yyy}^{\ast}$ in case II, $%
\Delta_{2xx,2yy}^{\ast}$ in case III, and $\Delta_{(3,1)y}^{\ast}$
in case IV are much larger than the values in case I. This is
because that the relaxation time recovering to balance is
inversely proportional to $s_{i}$. As $s_{i}$ decreases, the
corresponding mode will take more time to restore equilibrium, and
the deviation degree from the equilibrium increases. Physically,
the viscosity and heat conductivity of the physical system in case
II, the viscosity in case III, and the heat conductivity in case
IV are larger than the values in case I, which increase the
nonequilibrium behaviors of system.

2) $\Delta_{(3,1)x,(3,1)y,3xxx,3xxy,3xyy,3yyy}^{\ast}$ in case III
are similar to the values of case I,
$\Delta_{2xx,2yy,3xxy,3yyy}^{\ast}$ in case IV are smaller than
the values of case I. It can be explained as follow. The
relaxation parameters $s_{i}$ ($i=8,9,10,11,12,13$), density
gradient and temperature gradient in case III are consistent with
case I. The relaxation parameters $s_{i}$ ($i=5,7,11,13$) in case
IV are the same as case I, but the larger heat conductivity leads
to a decrease in density gradient and temperature gradient, which
reduce the nonequilibrium effect. There is a competition between
the viscosity, heat conduction and the gradient of physical
quantities.

3) $\Delta_{2xy,(3,1)x,3xxx,3xyy}^{\ast}$ in case I and III are
equal to zero, but the values in case II and IV are not equal to
zero. The reason is that, there is neither shear effect nor energy
flux in $x$ direction in case I and III ($u_x=0$), so
$\Delta_{2xy,(3,1)x,3xxx,3xyy}^{\ast}=0$. In case II and IV, it's
the opposite.

Figure 10 shows the viscosity, heat conductivity and Prandtl
number effects on the global average non-equilibrium
characteristics, (a) $Pr=0.5$, (b) $Pr=1.0$, (c) $Pr=2.0$. With
the increase of viscosity and heat conduction, $D_{TNE}$, $D_{2}$,
and $D_{(3,1)}$ will increase. The change of TNE strength is more
significant when heat conduction changes.  The growth of
$D_{2}$ and $D_{(3,1)}$ depend on the viscosity and thermal
conductivity, respectively. This further proves the correspondence
between $\Delta_{2\alpha \beta}^{\ast }$ and the viscosity term,
and the correspondence between $\Delta _{(3,1)\alpha }^{\ast }$
and the heat conduction term in NS equation. When the spike
arrives at the bottom of the calculation domain, or the RT
instability develops into the turbulent mixing stage, the global
average TNE strength and NOEF strength begin to decrease, and the
global average NOMF strength growth is slowing. The inclined
dashed lines roughly show the time that spikes reach the bottom
boundary of the calculation domain. When the viscosity and heat
conduction are relatively small, the spike develops relatively
quickly and reaches the bottom earlier. This is consistent with
the previous conclusion.
\begin{figure}[tbp]
\center\includegraphics*
[bbllx=14pt,bblly=14pt,bburx=315pt,bbury=230pt,width=0.8\textwidth]{%
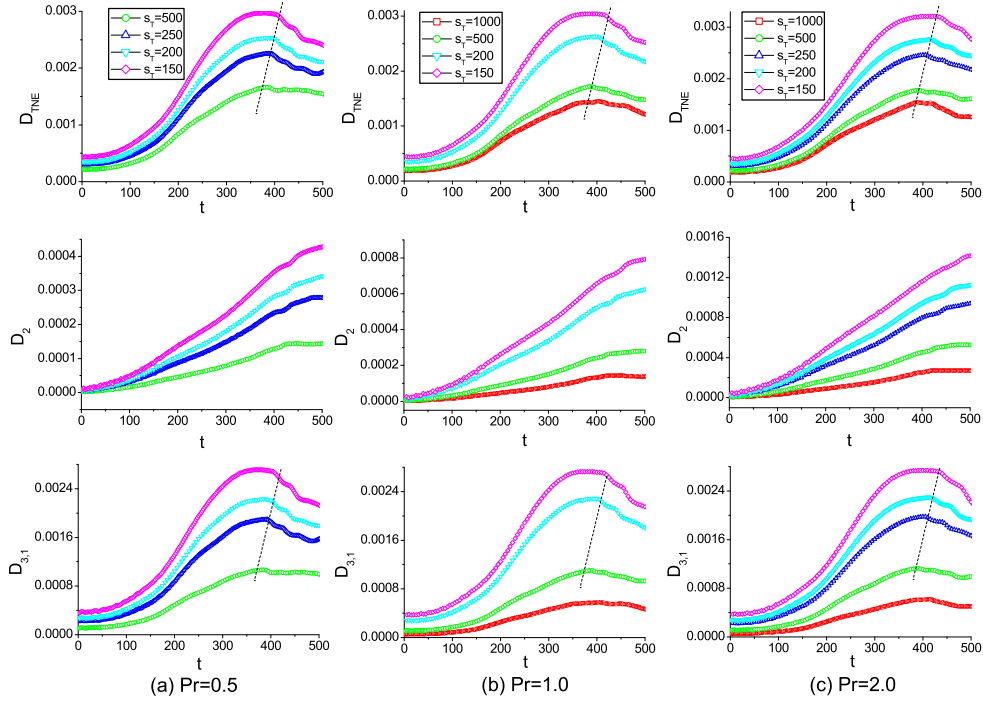} \caption{Prandtl number effects on the global average
non-equilibrium characteristics, (a) Pr=0.5, (b) Pr=1.0, (c)
Pr=2.0.}
\end{figure}

Figure 11 shows the snapshots of density non-uniformity $\delta
\rho$ and TNE strength $d$ at time $t=200$ and $t=400$. $\delta
\rho$ and $d$ demonstrate the HNE and TNE behaviours of the
system, respectively. In the position far from the perturbation
interface, $\delta \rho$ and $d$ are basically $0$. Around the
interface, particles with different density mix with each other,
and the exchanges of kinetic energy and momentum are produced,
$\delta \rho$ and $d$ are greater than zero. The characteristics
of density non-uniformity $\delta \rho$ and TNE strength $d$ are
quite consistent. HNE and TNE are 'the two-sides of a coin'. In
addition, both $\delta \rho$ and $d$ can be used to capture the
interface.

Fig. 12 shows the correlation degrees between macroscopic
non-uniformities and various global average nonequilibrium
strength in the case of $s_{v}=300$, $s_{T}=150$. In the figure,
considerably higher correlation degrees are founded between
density non-uniformity and the global average TNE strength
$D_{TNE}$, temperature non-uniformity and the global average NOEF
strength $D_{(3,1)}$, which are approximate to $1$. The
correlation degree between the velocity non-uniformity and the
global average NOMF strength $D_{2}$ is higher than that with
other non equilibrium strength.

\begin{figure}[tbp]
\center\includegraphics*
[bbllx=10pt,bblly=180pt,bburx=585pt,bbury=575pt,width=0.65\textwidth]{%
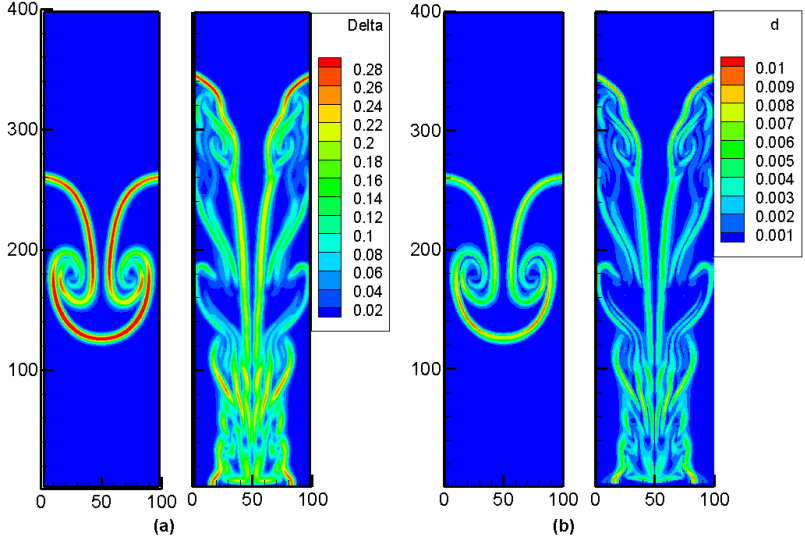} \caption{Snapshots of density non-uniformity $\delta
\rho$ (a) and TNE strength $d$ (b) at time $t=200$ and $t=400$.}
\end{figure}
\begin{figure}[tbp]
\center\includegraphics*
[bbllx=10pt,bblly=18pt,bburx=315pt,bbury=185pt,width=0.65\textwidth]{%
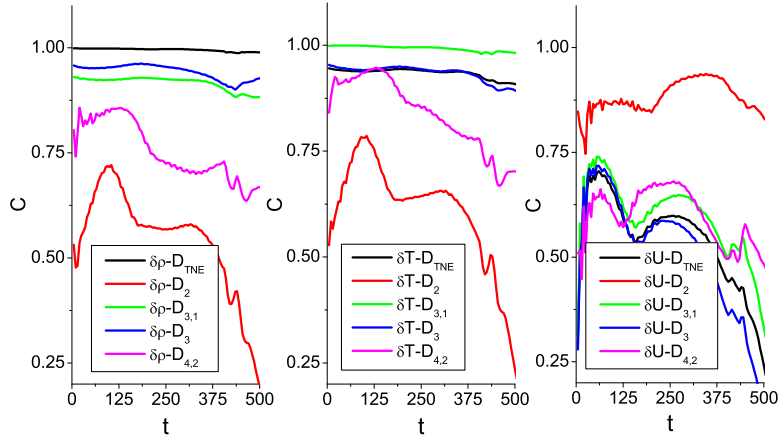} \caption{Correlation degrees between the macroscopic
non-uniformities and various global average nonequilibrium
strength. $\delta \rho$, $\delta T$ and $\delta U$ are density
non-uniformity, temperature non-uniformity and velocity
non-uniformity, respectively.}
\end{figure}
\begin{figure}[tbp]
\center\includegraphics*
[bbllx=15pt,bblly=15pt,bburx=315pt,bbury=235pt,width=0.75\textwidth]{%
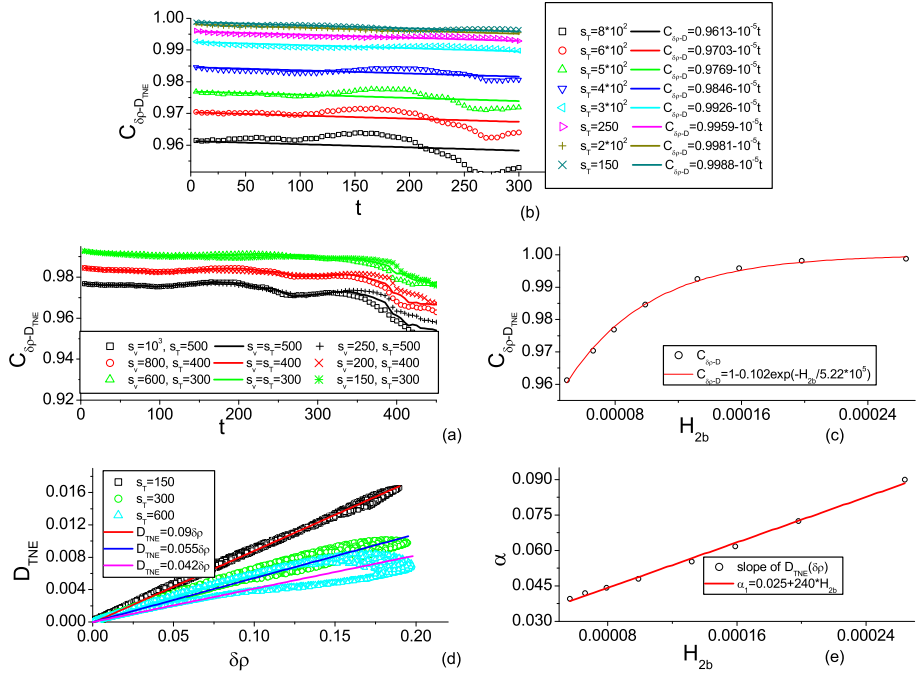} \caption{The correlation degree between $\delta \rho$
and $D_{TNE}$, (a) the effects of viscosity and heat conduction,
(b) (c) the variation of correlation degree with heat conduction,
(d) the linear relationship between $\delta \rho$ and $D_{TNE}$,
(e) the slope $\alpha$ of the linear relationship.}
\end{figure}
\begin{figure}[tbp]
\center\includegraphics*
[bbllx=0pt,bblly=20pt,bburx=580pt,bbury=450pt,width=0.65\textwidth]{%
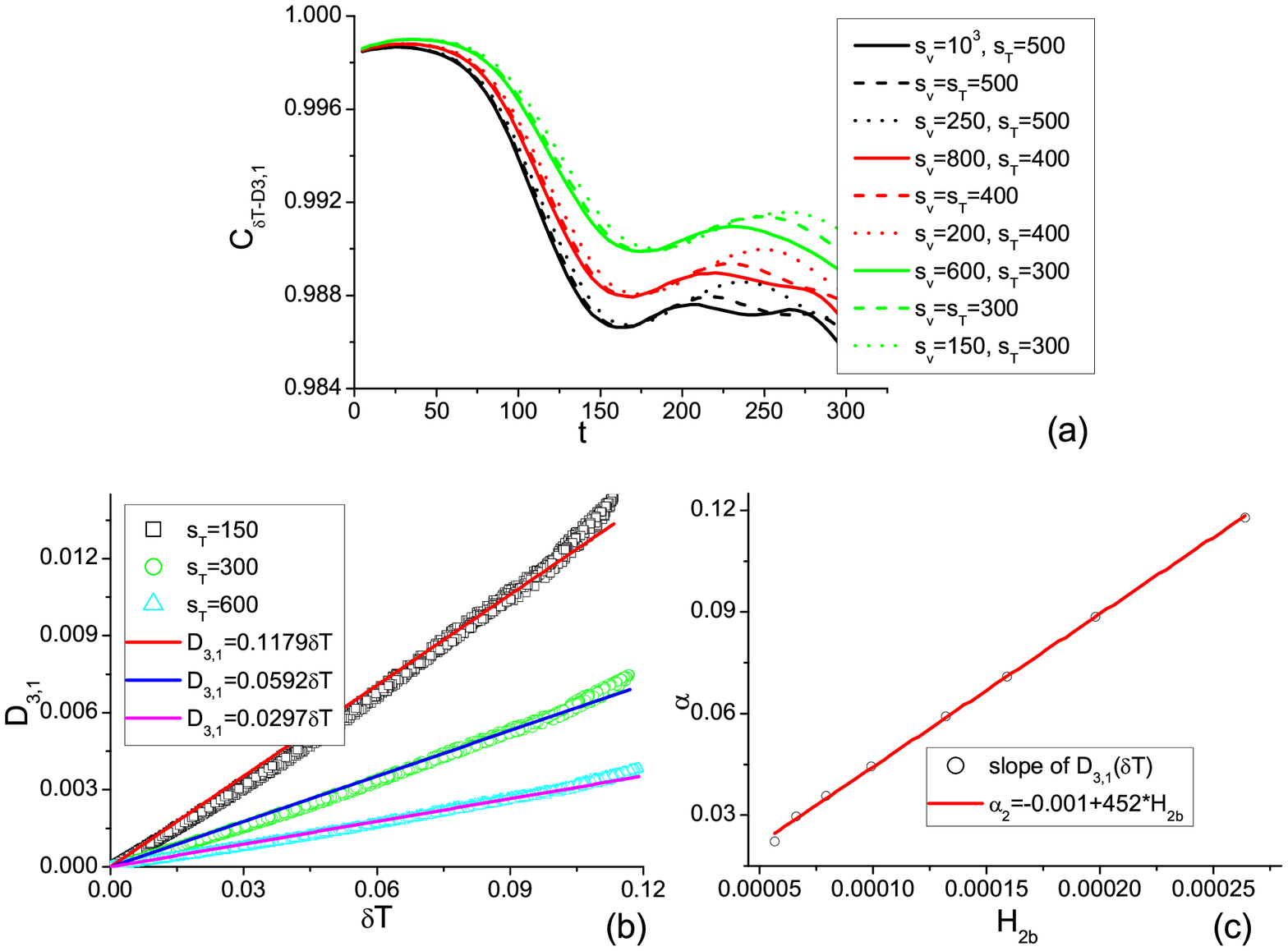} \caption{The correlation degree between $\delta T$ and
$D_{(3,1)}$, (a) the effect of viscosity and heat conduction,
(b)the linear relationship between $\delta T$ and $D_{(3,1)}$,
(c)the slope $\alpha$ of the linear relationship.}
\end{figure}

In Fig. 13(a) , we can find, the correlation degree between
$\delta \rho$ and $D_{TNE}$ varies with the viscosity and heat
conduction. Before the turbulent mixing stage, heat conduction
plays a major role. The greater the heat conduction, the higher
the degree of correlation. With the increase of heat conduction,
the correlation degree gradually tends to 1. (Fig. 13b). The trend
can be expressed by a exponential decay function(Fig. 13c),
\begin{equation}
C_{\delta \rho-D_{TNE}}=1-0.102exp(-H_{2b}\times 10^5/5.22),
H_{2b}=\sqrt{gk}/s_{T},
\end{equation}
where $H_{2b}$ is a relative thermal conductivity, $k$ is wave
number. In the turbulent mixing stage, the effect of viscosity is
reflected. When the heat conduction is constant, the higher the
viscosity is, the higher the degree of correlation. When the
correlation degree between the function A and B is equal to 1,
there is a linear relationship between A and B, that is $B=\alpha
A+\beta$. Fig. 13d shows the linear relationship between $\delta
\rho$ and $D_{TNE}$. The solid lines are the fitted curves. As can
be seen in the figure, the slope $\alpha$ of the linear
relationship is determined by the heat conduction,
$\alpha_{1}=0.025+240\times H_{2b}$.

In Fig. 14 , we can find, the correlation degree between $\delta
T$ and global average NOEF strength $D_{(3,1)}$ also varies with
the viscosity and heat conduction. Before the time $t=200$, heat
conduction plays a major role. The greater the heat conduction,
the higher the degree of correlation. The effect of viscosity in
the nonlinear stage is more obvious than that in the linear stage.
A linear relationship between $\delta T $ and $D_{(3,1)}$ is also
found, and the slope is also determined by the heat conduction,
$\alpha_{2}=-0.001+452\times H_{2b}$.

\section{Conclusions}

With a MRT discrete Boltzmann model, two-dimensional
Rayleigh-Taylor instability with different viscosity, thermal
conductivity and Prandtl number are simulated. Both viscosity and
heat conduction show significant inhibitory effect on RT
instability, and the inhibition effect is mainly achieved by
inhibiting the development of Kelvin-Helmholtz instability in the
reacceleration stage. Before this, the Prandtl number effect is
not sensitive. The non-equilibrium characteristics of system are
mainly probed. With the increase of viscosity or heat conduction,
different non-equilibrium components increase. There is a
competition between the viscosity, the heat conduction and the
gradient of physical quantities. When the RT instability develops
into the turbulent mixing stage, the global average TNE strength
and NOEF strength have a decrease. Correlation degrees between
macroscopic non-uniformities and various global average
nonequilibrium strength are analyzed. The correlation degrees
between density non-uniformity and the global average TNE
strength, temperature non-uniformity and the global average NOEF
strength, are approximate to $1$. Heat conduction shows a major
role on the correlation degree.

\section*{Acknowledgements}

FC acknowledges support of National Natural Science Foundation of
China [under Grant Nos. 11402138]. AX and GZ acknowledge support
of Foundation of LCP and National Natural Science Foundation of
China (under Grant No. 11475028).

\begin{appendix}

\section{CE expansion for the MRT DBM with gravity}

Using the Chapman-Enscog expansion on the two sides of discrete
Boltzmann equation, the Navier--Stokes equations with gravity term
can be derived.

We define
\begin{subequations}
\begin{equation}
\frac{\partial f_{i}}{\partial t}+v_{i\alpha }\frac{\partial
f_{i}}{\partial
x_{\alpha }}=-\mathbf{S}_{il}\left( f_{l}-f_{l}^{eq}\right) -f_{i}^{F}\text{,%
}  \label{ce1a}
\end{equation}%
\begin{equation}
f_{i}=f_{i}^{(0)}+f_{i}^{(1)}+f_{i}^{(2)}\text{,}  \label{ce1b}
\end{equation}%
\begin{equation}
\frac{\partial }{\partial t}=\frac{\partial }{\partial
t_{1}}+\frac{\partial }{\partial t_{2}}\text{,}  \label{ce1c}
\end{equation}%
\begin{equation}
\frac{\partial }{\partial x}=\frac{\partial }{\partial
x_{1}}\text{,} \label{ce1d}
\end{equation}%
where $f_{i}^{F}=g_{\alpha }\frac{(v_{i\alpha }-u_{\alpha
})}{RT}f_{i}^{eq}$, non-equilibrium parts $f_{i}^{(l)}=O(\epsilon
^{l})$, and the partial derivatives $\partial /\partial
t_{l}=O(\epsilon ^{l})$, $\partial /\partial x_{l}=O(\epsilon
^{l})$, $(l=1,2,\cdots )$. Equating the coefficients of the
zeroth, the first, and the second order terms in $\epsilon $ gives
\end{subequations}
\begin{subequations}
\begin{equation}
f_{i}^{(0)}=f_{i}^{eq}\text{,}  \label{ce2a}
\end{equation}%
\begin{equation}
(\frac{\partial }{\partial t_{1}}+v_{i\alpha }\frac{\partial
}{\partial x_{1\alpha
}})f_{i}^{(0)}=-\mathbf{S}_{il}f_{l}^{(1)}-f_{i}^{F}\text{,}
\label{ce2b}
\end{equation}%
\begin{equation}
\frac{\partial }{\partial t_{2}}f_{i}^{(0)}+(\frac{\partial }{\partial t_{1}}%
+v_{i\alpha }\frac{\partial }{\partial x_{1\alpha }})f_{i}^{(1)}=-\mathbf{S}%
_{il}f_{l}^{(2)}\text{.}  \label{ce2c}
\end{equation}%
They can be converted into moment space to obtain:
\end{subequations}
\begin{subequations}
\begin{equation}
\hat{\mathbf{f}}^{(0)}=\hat{\mathbf{f}}^{eq}\text{,}  \label{ce3a}
\end{equation}%
\begin{equation}
(\frac{\partial }{\partial t_{1}}+\hat{\mathbf{E}}_{\alpha }\frac{\partial }{%
\partial x_{1\alpha }})\hat{\mathbf{f}}^{(0)}=-\hat{\mathbf{S}}\hat{\mathbf{f%
}}^{(1)}-\mathbf{\hat{f}}^{F}\text{,}  \label{ce3b}
\end{equation}%
\begin{equation}
\frac{\partial }{\partial t_{2}}\hat{\mathbf{f}}^{(0)}+(\frac{\partial }{%
\partial t_{1}}+\hat{\mathbf{E}}_{\alpha }\frac{\partial }{\partial
x_{1\alpha }})\hat{\mathbf{f}}^{(1)}=-\hat{\mathbf{S}}\hat{\mathbf{f}}^{(2)}%
\text{,}  \label{ce3c}
\end{equation}%
where $\hat{\mathbf{E}}_{\alpha }=\mathbf{M}(v_{i\alpha }\mathbf{I})\mathbf{M%
}^{-1}$.

From Eq.\eqref{ce3b} we obtain
\end{subequations}
\begin{subequations}
\begin{equation}
\frac{\partial }{\partial t_{1}}\hat{f}_{1}^{eq}+\frac{\partial
}{\partial
x_{1}}\hat{f}_{2}^{eq}+\frac{\partial }{\partial y_{1}}\hat{f}_{3}^{eq}=-%
\hat{f}_{1}^{F}\text{,}  \label{ce4a}
\end{equation}%
\begin{equation}
\frac{\partial }{\partial t_{1}}\hat{f}_{2}^{eq}+\frac{\partial
}{\partial
x_{1}}\hat{f}_{5}^{eq}+\frac{\partial }{\partial y_{1}}\hat{f}_{6}^{eq}=-%
\hat{f}_{2}^{F}\text{,}  \label{ce4b}
\end{equation}%
\begin{equation}
\frac{\partial }{\partial t_{1}}\hat{f}_{3}^{eq}+\frac{\partial
}{\partial
x_{1}}\hat{f}_{6}^{eq}+\frac{\partial }{\partial y_{1}}\hat{f}_{7}^{eq}=-%
\hat{f}_{3}^{F}\text{,}  \label{ce4c}
\end{equation}%
\begin{equation}
\frac{\partial }{\partial t_{1}}\hat{f}_{4}^{eq}+\frac{\partial
}{\partial
x_{1}}\hat{f}_{8}^{eq}+\frac{\partial }{\partial y_{1}}\hat{f}_{9}^{eq}=-%
\hat{f}_{4}^{F}\text{,}  \label{ce4d}
\end{equation}%
\begin{equation}
\frac{\partial }{\partial t_{1}}\hat{f}_{5}^{eq}+\frac{\partial
}{\partial
x_{1}}\hat{f}_{10}^{eq}+\frac{\partial }{\partial y_{1}}\hat{f}%
_{11}^{eq}=-s_{5}\hat{f}_{5}^{(1)}-\hat{f}_{5}^{F}\text{,}
\label{ce4e}
\end{equation}%
\begin{equation}
\frac{\partial }{\partial t_{1}}\hat{f}_{6}^{eq}+\frac{\partial
}{\partial
x_{1}}\hat{f}_{11}^{eq}+\frac{\partial }{\partial y_{1}}\hat{f}%
_{12}^{eq}=-s_{6}\hat{f}_{6}^{(1)}-\hat{f}_{6}^{F}\text{,}
\label{ce4f}
\end{equation}%
\begin{equation}
\frac{\partial }{\partial t_{1}}\hat{f}_{7}^{eq}+\frac{\partial
}{\partial
x_{1}}\hat{f}_{12}^{eq}+\frac{\partial }{\partial y_{1}}\hat{f}%
_{13}^{eq}=-s_{7}\hat{f}_{7}^{(1)}-\hat{f}_{7}^{F}\text{,}
\label{ce4g}
\end{equation}%
\begin{equation}
\frac{\partial }{\partial t_{1}}\hat{f}_{8}^{eq}+\frac{\partial
}{\partial
x_{1}}\hat{f}_{14}^{eq}+\frac{\partial }{\partial y_{1}}\hat{f}%
_{15}^{eq}=-s_{8}\hat{f}_{8}^{(1)}-\hat{f}_{8}^{F}\text{,}
\label{ce4h}
\end{equation}%
\begin{equation}
\frac{\partial }{\partial t_{1}}\hat{f}_{9}^{eq}+\frac{\partial
}{\partial
x_{1}}\hat{f}_{15}^{eq}+\frac{\partial }{\partial y_{1}}\hat{f}%
_{16}^{eq}=-s_{9}\hat{f}_{9}^{(1)}-\hat{f}_{9}^{F}\text{.}
\label{ce4i}
\end{equation}%
From Eq.\eqref{ce3c} we obtain
\end{subequations}
\begin{subequations}
\begin{equation}
\frac{\partial }{\partial t_{2}}\hat{f}_{1}^{eq}=0\text{,}
\label{ce5a}
\end{equation}%
\begin{equation}
\frac{\partial }{\partial t_{2}}\hat{f}_{2}^{eq}+\frac{\partial
}{\partial
x_{1}}\hat{f}_{5}^{(1)}+\frac{\partial }{\partial y_{1}}\hat{f}_{6}^{(1)}=0%
\text{,}  \label{ce5b}
\end{equation}%
\begin{equation}
\frac{\partial }{\partial t_{2}}\hat{f}_{3}^{eq}+\frac{\partial
}{\partial
x_{1}}\hat{f}_{6}^{(1)}+\frac{\partial }{\partial y_{1}}\hat{f}_{7}^{(1)}=0%
\text{,}  \label{ce5c}
\end{equation}%
\begin{equation}
\frac{\partial }{\partial t_{2}}\hat{f}_{4}^{eq}+\frac{\partial
}{\partial
x_{1}}\hat{f}_{8}^{(1)}+\frac{\partial }{\partial y_{1}}\hat{f}_{9}^{(1)}=0%
\text{.}  \label{ce5d}
\end{equation}%
Adding Eqs.\eqref{ce4a}-\eqref{ce4d} and \eqref{ce5a}-\eqref{ce5d}
leads to the following equations,
\end{subequations}
\begin{subequations}
\begin{equation}
\frac{\partial }{\partial t}\hat{f}_{1}^{eq}+\frac{\partial }{\partial x}%
\hat{f}_{2}^{eq}+\frac{\partial }{\partial y}\hat{f}_{3}^{eq}=-\hat{f}%
_{1}^{F}\text{,}  \label{ce6a}
\end{equation}%
\begin{equation}
\frac{\partial }{\partial t}\hat{f}_{2}^{eq}+\frac{\partial }{\partial x}%
\hat{f}_{5}^{eq}+\frac{\partial }{\partial y}\hat{f}_{6}^{eq}=-\hat{f}%
_{2}^{F}-\frac{\partial }{\partial x}\hat{f}_{5}^{(1)}-\frac{\partial }{%
\partial y}\hat{f}_{6}^{(1)}\text{,}  \label{ce6b}
\end{equation}%
\begin{equation}
\frac{\partial }{\partial t}\hat{f}_{3}^{eq}+\frac{\partial }{\partial x}%
\hat{f}_{6}^{eq}+\frac{\partial }{\partial y}\hat{f}_{7}^{eq}=-\hat{f}%
_{3}^{F}-\frac{\partial }{\partial x}\hat{f}_{6}^{(1)}-\frac{\partial }{%
\partial y}\hat{f}_{7}^{(1)}\text{,}  \label{ce6c}
\end{equation}%
\begin{equation}
\frac{\partial }{\partial t}\hat{f}_{4}^{eq}+\frac{\partial }{\partial x}%
\hat{f}_{8}^{eq}+\frac{\partial }{\partial y}\hat{f}_{9}^{eq}=-\hat{f}%
_{4}^{F}-\frac{\partial }{\partial x}\hat{f}_{8}^{(1)}-\frac{\partial }{%
\partial y}\hat{f}_{9}^{(1)}\text{.}  \label{ce6d}
\end{equation}
\end{subequations}

It is easily shown that function $f_{i}^{F}$ satisfies the similar
moments.
\begin{subequations}
\begin{equation}
\int \int f^{F}d\mathbf{v}d\eta =0=\sum f_{i}^{F},  \label{ce7a}
\end{equation}%
\begin{equation}
\int \int f^{F}v_{\alpha }d\mathbf{v}d\eta =\rho g_{\alpha }=\sum
f_{i}^{F}v_{i\alpha },  \label{ce7b}
\end{equation}%
\begin{equation}
\int \int f^{F}\frac{\left( v^{2}+\eta ^{2}\right)
}{2}d\mathbf{v}d\eta =\rho g_{\alpha }u_{\alpha }=\sum
f_{i}^{F}\frac{\left( v_{i}^{2}+\eta _{i}^{2}\right) }{2},
\label{ce7c}
\end{equation}%
\begin{equation}
\int \int f^{F}v_{\alpha }v_{\beta }d\mathbf{v}d\eta =\rho
g_{\alpha }u_{\beta }+\rho g_{\beta }u_{\alpha }=\sum
f_{i}^{F}v_{i\alpha }v_{i\beta }, \label{ce7d}
\end{equation}%
\begin{eqnarray}
\int \int f^{F}\frac{\left( v^{2}+\eta ^{2}\right) }{2}v_{\alpha }d\mathbf{v}%
d\eta  &=&\rho \lbrack g_{\beta }u_{\alpha }u_{\beta }+(\frac{b+2}{2}RT+%
\frac{u^{2}}{2})g_{\alpha }]  \notag \\
&=&\sum f_{i}^{F}\frac{\left( v_{i}^{2}+\eta _{i}^{2}\right)
}{2}v_{i\alpha }.  \label{ce7e}
\end{eqnarray}
\end{subequations}
Eqs.\eqref{ce7a}-\eqref{ce7e} can be written in a matrix form,
i.e., $\mathbf{\hat{f}}^{F}= \mathbf{Mf}^{F}$, where
$\hat{f}_{1}^{F}=0$, $\hat{f}_{2}^{F}=\rho g_{x}$,
$\hat{f}_{3}^{F}=\rho
g_{y} $, $\hat{f}_{4}^{F}=\rho (g_{x}u_{x}+g_{y}u_{y})$, $\hat{f}%
_{5}^{F}=2\rho g_{x}u_{x}$, $\hat{f}_{6}^{F}=\rho (g_{x}u_{y}+g_{y}u_{x})$, $%
\hat{f}_{7}^{F}=2\rho g_{y}u_{y}$, $\hat{f}_{8}^{F}=\rho \lbrack
g_{x}u_{x}^{2}+g_{y}u_{x}u_{y}+g_{x}(\frac{b+2}{2}RT+\frac{u^{2}}{2})]$, $%
\hat{f}_{9}^{F}=\rho \lbrack g_{y}u_{y}^{2}+g_{x}u_{x}u_{y}+g_{y}(\frac{b+2}{%
2}RT+\frac{u^{2}}{2})]$, and the others ($i=10$, \ldots , $16$)
are $0$.

Using the definitions of $\hat{f}_{i}^{eq}$ and $\hat{f}_{i}^{F}$,
we can obtain:
\begin{subequations}
\begin{equation}
\frac{\partial \rho }{\partial t}+\frac{\partial (\rho u_{\alpha })}{%
\partial x_{\alpha }}=0\text{,}  \label{ce8a}
\end{equation}%
\begin{equation}
\frac{\partial (\rho u_{\alpha })}{\partial t}+\frac{\partial
\left( \rho
u_{\alpha }u_{\beta }\right) }{\partial x_{\beta }}+\frac{\partial P}{%
\partial x_{\alpha }}=\frac{\partial }{\partial x_{\beta }}[\mu (\frac{%
\partial u_{\alpha }}{\partial x_{\beta }}+\frac{\partial u_{\beta }}{%
\partial x_{\alpha }}-\frac{2}{b}\frac{\partial u_{\chi }}{\partial x_{\chi }%
}\delta _{\alpha \beta })]-\rho g_{\alpha }\text{,} \label{ce8b}
\end{equation}%
\begin{eqnarray}
\frac{\partial e}{\partial t}+\frac{\partial }{\partial x_{\alpha
}}\left[
(e+P)u_{\alpha }\right] &=&\frac{\partial }{\partial x_{\beta }}[(\frac{b}{2}%
+1)\lambda ^{\prime }R\frac{\partial T}{\partial x_{\beta
}}+\lambda ^{\prime }(\frac{\partial u_{\alpha }}{\partial
x_{\beta }}+\frac{\partial
u_{\beta }}{\partial x_{\alpha }}-\frac{2}{b}\frac{\partial u_{\chi }}{%
\partial x_{\chi }}\delta _{\alpha \beta })u_{\alpha }]-\rho g_{\alpha
}u_{\alpha }\text{,} \notag\\
(\alpha ,\beta ,\chi &=&x,y).\label{ce8c}
\end{eqnarray}
\end{subequations}
where $\mu =\rho RT/s_{v}$, ($s_{v}=s_{5}=s_{6}=s_{7}$), $\lambda
^{\prime }=\rho RT/s_{T}$, ($s_{T}=s_{8}=s_{9}$).

By modifying the collision operators of the moments related to
energy flux:
\begin{subequations}
\begin{align}
& \hat{\mathbf{S}}_{88}(\hat{f}_{8}-\hat{f}_{8}^{eq})\Rightarrow \hat{%
\mathbf{S}}_{88}(\hat{f}_{8}-\hat{f}_{8}^{eq})+(s_{T}/s_{v}-1)\rho
Tu_{x}
\notag \\
& \qquad \times (2\frac{\partial u_{x}}{\partial x}-\frac{2}{b}\frac{%
\partial u_{x}}{\partial x}-\frac{2}{b}\frac{\partial u_{y}}{\partial y}%
)+(s_{T}/s_{v}-1)\rho Tu_{y}(\frac{\partial u_{y}}{\partial x}+\frac{%
\partial u_{x}}{\partial y})\text{,}
\end{align}%
\begin{align}
& \hat{\mathbf{S}}_{99}(\hat{f}_{9}-\hat{f}_{9}^{eq})\Rightarrow \hat{%
\mathbf{S}}_{99}(\hat{f}_{9}-\hat{f}_{9}^{eq})+(s_{T}/s_{v}-1)\rho
Tu_{x}
\notag \\
& \qquad \times (\frac{\partial u_{y}}{\partial x}+\frac{\partial u_{x}}{%
\partial y})+(s_{T}/s_{v}-1)\rho Tu_{y}(2\frac{\partial u_{y}}{\partial y}-%
\frac{2}{b}\frac{\partial u_{x}}{\partial x}-\frac{2}{b}\frac{\partial u_{y}%
}{\partial y})\text{,}
\end{align}%
we get the following energy equation:
\end{subequations}
\begin{equation}
\frac{\partial e}{\partial t}+\frac{\partial }{\partial x_{\alpha
}}\left[
(e+P)u_{\alpha }\right] =\frac{\partial }{\partial x_{\beta }}[\lambda \frac{%
\partial T}{\partial x_{\beta }}+\mu (\frac{\partial u_{\alpha }}{\partial
x_{\beta }}+\frac{\partial u_{\beta }}{\partial x_{\alpha }}-\frac{2}{b}%
\frac{\partial u_{\chi }}{\partial x_{\chi }}\delta _{\alpha \beta
})u_{\alpha }]-\rho g_{\alpha }u_{\alpha }\text{,}
\end{equation}
where $\lambda =(\frac{b}{2}+1)R\lambda ^{\prime }$.
\end{appendix}


\end{document}